\documentclass[5p,twocolumn]{elsarticle}

\usepackage{graphics}
\usepackage{graphicx}
\usepackage{epsfig}

\usepackage{amssymb,amsmath}
\usepackage[latin1]{inputenc}
\newcommand{\YBCO}{Y$_{1.6}$Ba$_{2.3}$Cu$_{3.3}$O$_{7-x}$}
\newcommand{\BO}{B$_2$O$_3$}

\begin{document}

\begin{frontmatter}


%\corauth[cor1]{}
%\corauthref[cor1]{Tel.:+39 0916234207; fax: +39 0916162461}

\title{Effect of boron doping in the microwave surface resistance of neutron irradiated melt-textured \YBCO\ samples}

\author[label1]{A. Agliolo Gallitto}
\author[label1]{M. La Duca}
\author[label1,cor1]{M. Li Vigni}
\author[label2]{U. Topal}
\author[label3]{\c{S}. Yildiz}
\address[label1]{CNISM and Dipartimento di Fisica, Universit\`{a} di Palermo, via Archirafi 36, 90123 Palermo, Italy}
\address[label2]{TUBITAK-UME (National Metrology Institute), P. K. 54, 41470 Gebze-Kocaeli, Turkey}
\address[label3]{Department of Physics, Faculty of Science and Arts, Gaziosmanpasa University, 60240 Tokat, Turkey}
%\corauth[cor1]{Tel.:+39 0916234208; fax: +39 0916162461; e-mail: livigni@fisica.unipa.it}
%\ead[ur1]{livigni@fisica.unipa.it}

\begin{abstract}
We report on the microwave surface resistance of melt-textured \YBCO\ samples, doped with different amount of \BO\ and, subsequently, irradiated by thermal neutrons at the fluence of $1.476\times 10^{17}~\mathrm{cm^{-2}}$. The microwave surface resistance has been measured as a function of temperature and DC magnetic field. The experimental results are quantitatively discussed in the framework of the Coffey and Clem theory, properly adapted to take into account the d-wave nature of cuprate superconductors. By fitting the experimental data at zero DC field, we have highlighted the effects of the induced defects in the general properties of the samples, including the intergranular region. The analysis of the results obtained at high DC fields allowed us to investigate the fluxon dynamics and deduce the depinning frequency; in particular, we have shown that the addition of \BO\ up to 0.1~wt\% increases the effectiveness of the defects to hinder the fluxon motion induced by the microwave current.      
\end{abstract}

\begin{keyword}
% Here, in the form: keyword \sep keyword
Depinning frequency \sep Microwave surface resistance \sep Thermal-neutron irradiation \sep \BO\ addition

\PACS 74.25.Nf \sep 74.60.Ge \sep 74.62.Dh
%
%74.25.Nf(Response to electromagnetic fields (nuclear magnetic resonance, surface impedance, etc.)), 74.60.Ge (Flux pinning, flux creep, and flux-line lattice dynamics), 74.62.Dh (Effects of crystal defects, doping and substitution)
\end{keyword}

\end{frontmatter}

% main text
\section{Introduction}
\label{sec:introduz}

Since the discovery of high-$T_c$ superconductors (HTS), different methods have been exploited to introduce pinning centers that, hindering the fluxon motion, lower the energy losses. Two different methods can be used to insert defects acting as pinning centers: introduction of normal-phase impurities, during the synthesis process, and/or damage by irradiation. Many experiments have been performed to investigate the irradiation effects on electromagnetic properties of HTS. Irradiation with protons~\cite{APL91,LeeIEEE07} and heavy ions~\cite{Ghigo,GhigoMW,LOF_PRB51} involves charged particles that strongly interact with the solid matter, producing dislocations and/or columnar defects. Neutron irradiation may give rise to a more homogeneous distribution of defects because it involves uncharged particles, which can penetrate freely into the matter~\cite{Koler,SauPRB51,Akiyoshi,Kulikov}. Depending on the neutron energy, neutron irradiation creates clusters or cascades of point defects randomly distributed into the materials, whose dimensions range from few nm to 10 nm ~\cite{Akiyoshi,Eisterer,Tonies,5diTopal}. The irradiation-induced defects interact with the pre-irradiation defect structure either through the direct replacement of many point defects by a large collision cascade, or by statistical rearrangement of certain atoms (mostly oxygen atoms); so, the radiation effectiveness depends on the preexisting defect structure in the sample~\cite{Koler,PRB43_91,Stefanescu}. In order to enhance the interaction of incident neutrons with the superconducting matrix, cuprate superconductors have been doped with elements having large neutron-absorption cross sections, such as Li~\cite{Stefanescu,Zhao}, B~\cite{Topal2003,Topal2004} and U~\cite{Eisterer,Tonies,Sawh,Marinaro}, or prepared introducing normal-phase particles during the synthesis process~\cite{Topal2003,Topal2004,PhysicaB}.  

Topal et \emph{al.}~\cite{Topal2003,Topal2004,Topal2005} have investigated the effects of thermal-neutron irradiation and B doping in a series of melt-textured YBCO samples containing pre-irradiation defects due to inclusion in the YBCO-123 phase of a certain amount of non-superconducting YBCO-211 phase, present as particles of micrometric dimensions uniformly distributed over the sample. In order to introduce B atoms, they added B$_2$O$_3$ powder after calcinating the 211 and 123 phases. Investigating the unirradiated samples, it has been shown that the only addition of \BO\ enlarges the intergrain regions, weakening conduction links between grains and lowering both $T_c$ and the critical current density, $J_c$~\cite{Topal2005}. However, in the same paper, it has been highlighted a reduction of lattice parameters by 0.1-0.2\% with \BO\ addition up to 0.2-0.5 wt\%, suggesting that some B atoms substitute copper atoms, probably in the Cu-O chains. So, one can infer that part of B atoms resides at grain boundaries and the remaining enters in the matrix structure of the superconducting grains. This hypothesis is consistent with the results obtained by Margiani et \emph{al.}~\cite{Margiani} and Katsura et \emph{al.}~\cite{Katsura}. 

The effects of neutron irradiation on the above-mentioned B-doped samples has been investigated by measuring the field dependence of $J_c$~\cite{Topal2003,Topal2004}. The authors observe an increase of $J_c$ after irradiation more enhanced in the B-doped than in the undoped samples; moreover, the effectiveness of neutron irradiation increases on increasing the B$_2$O$_3$ content until a threshold value of \BO\ content is overtaken~\cite{Topal2003}, the  highest enhancement of $J_c$ being obtained in the sample with 0.1 wt\% of \BO. By investigating samples obtained with 10~mol\% and 30~mol\% of 211 phase, it has been found that $J_c$ does not depend on the 211 phase content in the unirradiated samples, while the increase of $J_c$ induced by neutron irradiation is more enhanced in the samples prepared with 30~mol\% of the 211 phase~\cite{Topal2004}. These findings strongly suggest that both the introduction of the 211 particles and the \BO\ addition make the neutron radiation more effective in creating pinning centers; moreover, they confirm the hypothesis that part of the B atoms enters in the superconducting grains.

In this paper, we investigate the microwave (mw) surface resistance, $R_s$, of a series of melt-testured \YBCO\ samples prepared starting from a mixture of 23\% YBCO-211 and 77\% YBCO-123 phases, subsequently doped with different amount of \BO\ and, eventually, irradiated by thermal neutrons at the fluence of $1.476\times 10^{17}~\mathrm{cm^{-2}}$. Details of the preparation method are reported in~\cite{Topal2003}. $R_s$ has been measured as a function of the temperature in the absence of DC magnetic field and as a function of the DC magnetic field at fixed temperatures. The experimental results are discussed in the framework of the Coffey and Clem theory~\cite{CC}, properly adapted to take into account the d-wave nature of cuprate superconductors. Firstly, we have investigated an undoped and unirradiated sample to check the analysis method; successively, we have used the same method to quantitatively  discuss the effects of \BO\ addition on the magnetic-field induced mw losses.

The mw response of irradiated YBCO samples, mainly films, has been investigated by several authors~\cite{APL91,LeeIEEE07,GhigoMW,LOF_PRB51,Stefanescu,APL90}. The studies can be listed in two categories: i) investigation as a function of the temperature at zero DC magnetic fields~\cite{APL91,LeeIEEE07,APL90}, in which an important role is played by the intergranular medium; ii) investigation of SC in the mixed state~\cite{GhigoMW,LOF_PRB51,Stefanescu,Powell}, which highlights properties related to flux pinning. It has been found that proton irradiation does not considerably affect the residual surface resistance at low temperatures, while gives rise to a decrease of $R_s$ at intermediate temperatures due to an enhanced impurity scattering rate~\cite{APL91,LeeIEEE07}. In the mixed state, columnar defects induced by heavy-ion irradiation effectively pin the flux lines, and reduce the nonlinear effects~\cite{GhigoMW,LOF_PRB51,Powell}. To the best of our knowledge, only a few papers deal with the mw response of neutron-irradiated HTS in the mixed state~\cite{Stefanescu,Chash}. Ref.~\cite{Stefanescu} reports on results of mw absorption in Li-doped YBCO sample; although the authors do not discuss the dynamics of fluxons, they highlight that the effects of neutron irradiation strongly depend on the pre-irradiation defect structure. 

In this paper, we devote the attention mainly to the investigation of the mw response in the mixed state since mw measurements allow one to conveniently investigate the vortex dynamics~\cite{golo}; indeed they probe the vortex response at very low currents, when vortices undergo reversible oscillations and are less sensitive to flux-creep processes. A very important parameter of vortex dynamics is the depinning frequency, $\omega_0$, defined as the ratio of the restoring-pinning-force coefficient and the flux-line viscosity~\cite{gittle}; so $\omega_0$ gives direct information of the pinning strength~\cite{GOLOSOVSKY,bonura}. Our results show that on increasing the \BO\ content the depinning frequency increases, demonstrating an increase of the pinning effectiveness. 

\section{Experimental apparatus and samples}\label{sec:samples}
\begin{figure}[h!]
  \centering
  \includegraphics[width=0.45\textwidth]{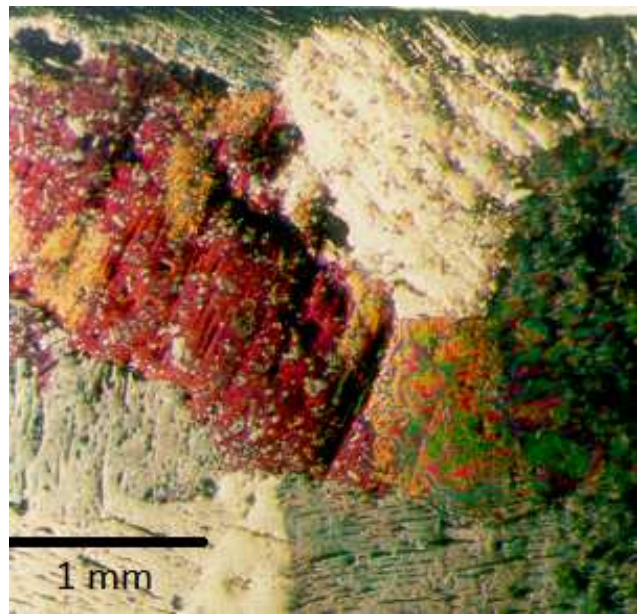}\\
  \caption{Polarizing microscope image of a portion of a specimen obtained after inclusion of YBCO-211 phase into YBCO-123, without doping and irradiation; inclusions are visible as dots, better visible in the red domain.}\label{Fig:foto}
\end{figure}

The mw surface resistance has been investigated in five melt-textured \YBCO\ samples, doped with different amount of \BO\ and, successively, irradiated in a TRIGA-MARK-II research reactor (energy ranging from few meV to 10 MeV). Irradiation has been performed with thermal neutrons ($\sim$25~meV) of flux density $8.2 \times10^{12}~\mathrm{cm^{-2} s^{-1}}$ at the corresponding fluence $1.476\times 10^{17}~\mathrm{cm^{-2}}$. The procedure for the sample preparation, doping and irradiation is reported in detail elsewhere~\cite{Topal2003}. The starting \YBCO\ material (unirradied and undoped) has been prepared to have a mixture of 23\% YBCO-211 and 77\% YBCO-123 phases. Figure~\ref{Fig:foto} shows a polarizing microscope image of a portion of a specimen obtained after calcination of the two phases; different colors indicate different domains. The small particles, which are better visible over the red domain, correspond to 211 inclusions; most of them have dimensions of 1-2 $\mu$m, except a small part extending up to nearly 10~$\mu$m, and are uniformly distributed over the specimen.  

After calcination, the starting material was powdered and divided in four parts; one part was kept boron free and the other parts were mixed with a certain amount of \BO\ powder. The powders were pressed into pellets, subjected to further thermal and annealing process~\cite{Topal2003} and, subsequently, irradiated with thermal neutrons at the same fluence. We have investigated five samples of parallelepiped shape, with approximate dimensions: $t\sim 1~\mathrm{mm}, w\sim 2~\mathrm{mm}, h\sim 3~\mathrm{mm}$; the one that we indicate as Y-00 is obtained from the starting material (without irradiation and doping); a second one (Y-10) is boron free and irradiated; the others, Y-11, Y-12 and Y-13, are obtained doping the starting material with  0.05, 0.1 and 0.5 wt\% of \BO, respectively, and successively irradiating them at the same fluence as sample Y-10. 

\begin{figure}[b]
\centering
\includegraphics[width=0.45\textwidth]{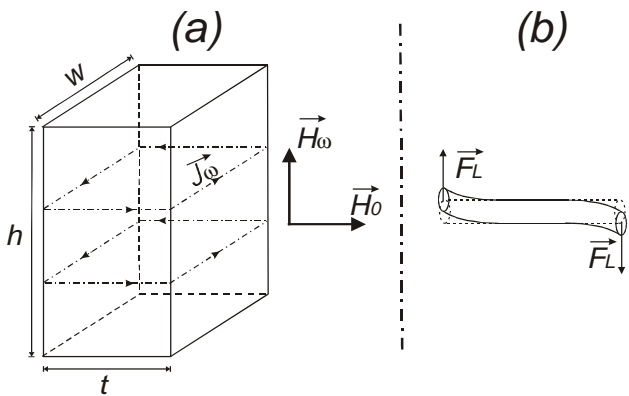}
\caption{(a) Field and current geometry at the sample surface. (b) Schematic representation of the motion of a flux line.}
\label{fig:sample}
\end{figure}

The mw surface resistance, $R_s$, has been measured using the cavity-perturbation technique~\cite{golo}. A copper cavity, of cylindrical shape with golden-plated walls, is tuned in the $\mathrm{TE}_{011}$ mode resonating at $\omega/2\pi \approx 9.6$~GHz. The sample is located in the center of the cavity by a sapphire rod, in the region in which the mw magnetic field is maximum. The cavity is placed between the poles of an electromagnet which generates DC magnetic fields up to $\mu_0H_0\approx 1$~T. Two additional coils, independently fed, allow compensating the residual
field and working at low magnetic fields. A cryostat and a temperature controller allow working either at fixed temperatures or at temperature varying with a constant rate. The sample and the field geometries are shown in Fig.~1a; the DC magnetic field, $\emph{\textbf{H}}_{0}$, is perpendicular to the mw magnetic field, $\emph{\textbf{H}}_{\omega}$. When the sample is in the mixed state, the induced mw current, $J_{\omega}$, causes a tilt motion of the vortex lattice~\cite{BRANDT} because of the Lorentz force, $F_L$; Fig.~\ref{fig:sample}b schematically shows the motion of a flux line.

The surface resistance of the sample, which is proportional to the mw energy losses, is given by\\
\begin{equation}
    R_s= \Gamma \left(\frac{1}{Q_L} - \frac{1}{Q_U}\right)\,,
\end{equation}
where $Q_L$ is the quality factor of the cavity loaded with the sample, $Q_U$ that of the empty cavity and $\Gamma$ the geometry factor of the sample.\\
The quality factor of the cavity has been measured by an hp-8719D Network Analyzer. All the measurements have been performed at very low input mw power; the estimated amplitude of the mw magnetic field in the region in which the sample is located is of the order of $0.1~\mu$T.

\section{Experimental results and discussion}\label{sec:results}
The surface resistance has been measured as a function of the temperature, in the absence of magnetic field, and as a function of the DC magnetic field, at fixed temperatures. We will present and discuss the temperature dependence and the magnetic field dependence separately. The results obtained at zero DC magnetic field allow us to obtain information on general properties of the whole sample, including the intergranular regions. The discussion of the results as a function of the DC magnetic field is focused to investigate the fluxon dynamics in the superconducting grains.  

\subsection{Surface resistance at zero magnetic field}
\begin{figure}[h]
\centering
\includegraphics[width=0.45\textwidth]{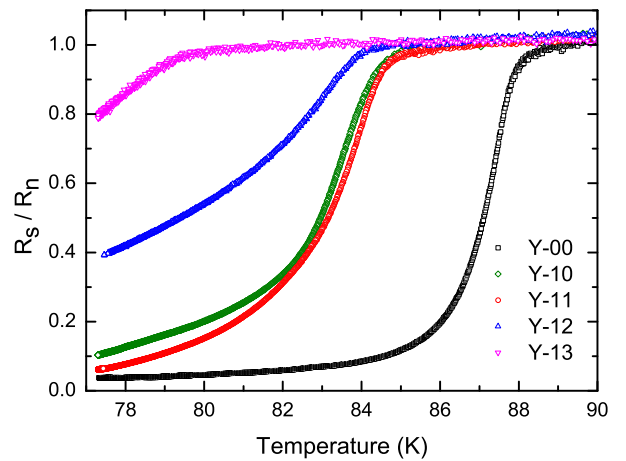}
\caption{Normalized values of the mw surface resistance, $R_s/R_n$, as a function of the temperature obtained at $H_0=0$ for the investigated YBCO samples: Y-00 is the undoped and unirradiated sample; Y-10 is undoped and irradiated at the neutron fluence of $1.476\times 10^{17}~\mathrm{cm^{-2}}$; Y-11, Y-12 and Y-13 are doped with 0.05, 0.1 and 0.5 wt\% of \BO, respectively, and subsequently irradiated at the same fluence as sample Y-10.}
\label{fig:Rs(T)}
\end{figure}

The temperature dependence of the mw surface resistance obtained in the five samples at $H_0=0$ is reported in Fig.~\ref{fig:Rs(T)} in the temperature range $77 \div 90$~K. In order to disregard the geometry factor and compare results obtained in samples of different, though similar, dimensions, we report $R_s(T)$ normalized to the near-$T_c$ normal-state value, $R_n$.

The results of Fig.~\ref{fig:Rs(T)} show that the irradiation in the B-free sample causes a $T_c$ reduction of about 3~K and a broadening of the $R_s(T)$ curve; moreover, looking in detail at the low-$T$ region, one can note a shoulder around 79~K that suggests the presence of a second superconducting phase at low $T$. Sample Y-11 exhibits nearly the same transition temperature as Y-10, but the $R_s(T)$ curve does not show visible structures, indicating that Y-11 sample is more homogeneous. This could be due to the fact that the presence of B atoms, having large neutron-absorption cross section, makes the irradiation effect more homogeneous over the sample. The increase of \BO\ content from 0.05 to 0.1 wt\% causes a reduction of $T_c$ of about 1~K, while a high \BO\ content (0.5 wt\% in Y-13) causes a $T_c$ drop of about 5~K and a large broadening (not fully visible here), which suggests a degradation of the superconducting properties of the YBCO material.

In the absence of DC magnetic fields, the variation with the temperature of the mw surface resistance is related to the temperature dependence of the quasiparticle density.
In the London local limit, the surface resistance is proportional to the imaginary part of the complex penetration depth of the electromagnetic field, $\widetilde{\lambda}$:
\begin{equation}\label{Rs}
    R_s=-\mu_{0}\omega ~\mathrm{Im}[{\widetilde{\lambda}(\omega,T)}].
\end{equation}
In the framework of the two-fluid model, $\widetilde{\lambda}$ is related to the temperature dependence of the London penetration depth and the normal skin depth
\begin{equation}\label{lambdat}
   \widetilde{\lambda}(\omega,T)=\left[\frac{1}{\lambda^{2}(T)}-\frac{2i}{\delta_{n}^{2}(\omega,T)}\right]^{-1/2}\,,
\end{equation}
with
\begin{equation}\label{lambda0}
\lambda^2(T) = \frac{\lambda_0^2}{1-w_0(T)}\,,
\end{equation}
\begin{equation}\label{delta0}
\delta_{n}^2(\omega,T) = \frac{\delta_0^2(\omega)}{w_0(T)}\,,
\end{equation}
where $\lambda_0$ is the London penetration depth at $T = 0$, $\delta_0$ is the normal-fluid skin depth at $T = T_c$, $w_0(T)$ is the fraction of normal electrons at $H_0 = 0$.\\

Since the normal-state value of the mw surface resistance is given by $R_n=\mu_0\omega\delta_0/2$, from Eqs.~(2-5) one can see that the expected temperature dependence of $R_s/R_n$, at $H_0=0$, depends on the ratio $\lambda_0/\delta_0$ and $w_0(T)$. Moreover, for inhomogeneous samples containing defects it is necessary to consider a residual surface resistance and a distribution of $T_c$ over the sample. In order to fit the experimental data, we have assumed $w_0(T)=(T/T_c)^2$ consistently with the results reported in the literature for the temperature dependence of the quasiparticle density in cuprate superconductors~\cite{anlage,Bonn,Porch}, which has been ascribed to the d-wave nature of these materials. Furthermore, we have averaged the expected curve over a Gaussian distribution function of $T_c$. In order not to include the residual surface resistance as additional parameter, we have fitted the data imposing that, for each sample, the expected curve coincides with the experimental one at the lowest temperature investigated; practically, this means to fit the temperature-induced variations of $R_s/R_n$. The only free fitting parameter is $\lambda_0/\delta_0$, though also $T_{c0}$ and $\sigma_{T_c}$ play a role near the onset of the superconducting transition. 

\begin{figure}[h]
\centering
\includegraphics[width=0.45\textwidth]{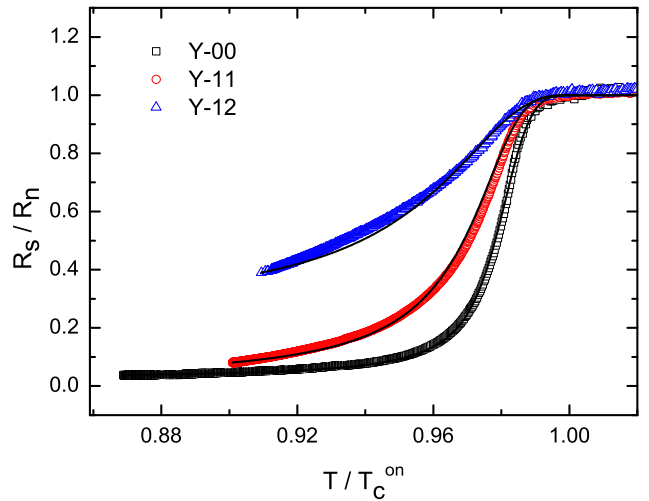}
\caption{$R_s/R_n$ as a function of the reduced temperature, $T/T_c^{on}$, at $H_0=0$ for the three YBCO samples: Y-00 (pristine), Y-11 and Y-12, along with the best-fit curves (continuous lines) obtained as explained in the text.}
\label{fig:Rs(T)FIT}
\end{figure}

Figure \ref{fig:Rs(T)FIT} shows the experimental data for the samples Y-00, Y-11 and Y-12 along with the best-fit curves obtained as above described; to better highlight the different behavior of $R_s/R_n$ in the different samples the results have been reported as a function of the reduced temperature, $T/T_c^{on}$, where $T_c^{on}$ is the temperature value at which $R_s/R_n=1$. The best-fit parameters used for the three samples are reported in Tab.~\ref{tab:samples}. We have tried to fit also the results obtained in sample Y-10; for this sample we have obtained a good fit only for $T \geqslant 83$~K, using parameters similar as those used for sample Y-11, while the low-$T$ behavior cannot be justified probably because of the presence of the low-$T$ phase. We have leaved out sample Y-13 because it is visibly degraded. 

Tab.~\ref{tab:samples} summarizes some properties of the five investigated samples and reports the parameters used to obtain the best-fit curves of Fig.~\ref{fig:Rs(T)FIT}; $R_n$ is the value of the mw surface resistance at $T=T_c^{on}$, which has been determined by Eq.~(1) considering the geometry factor of the samples. From the values of $R_n$, one can see that also in the normal state the sample with the highest \BO\ addition degrades, indeed we obtain an increase of about 60\% of $R_n$ corresponding to an increase of the normal-state resistivity of about 35\%. A comparison of the results reported in the table, for the samples for which the $R_s(T)$ curves have been fitted, shows that, besides a variation of the transition temperature, the main effect of the B content is to increase the ratio $\lambda_0/\delta_0$. 

From the estimated values of $R_n$, one can estimate the normal skin depth; in particular, we obtain $\delta_0\approx 60~\mu$m for samples Y-00, Y-10 and Y-11, and  $\delta_0\approx 80~\mu$m and $90~\mu$m for the samples Y-12 and Y-13, respectively. Although the values of $\delta_0$ are about the same in samples Y-00 and Y-11, the ratio $\lambda_0/\delta_0$ increases with the addition of \BO; a further increase occurs in sample Y-12. This finding is ascribable to an increase of the effective penetration depth, which in granular samples is ruled by the intergrain field penetration depth, and is consistent with the observation that part of B atoms resides at the grain boundaries reducing the intergrain Josephson current~\cite{Topal2005}. 

\begin{table}[t]
\centering \small
\begin{tabular}{ccccccc}
\hline
\textbf{Sample} & \BO\  & $T_c^{on}$ & $T_{c0}$ & $\sigma_{T_c}$ & $\lambda_0/\delta_0$ & $R_n$ \\
                & (wt\%)& [K] & [K]       &   [K]   &     & [$\mathrm{\Omega}$] \\
\hline
\textbf{Y-00}& 0    & 89   & 87.8 & 0.45 &0.089 & 2.3 \\
%\hline
\textbf{Y-10}& 0    & 85.7 & -- & -- &--& 2.2 \\
%\hline
\textbf{Y-11}& 0.05 & 85.8 & 84.6  & 0.45 & 0.135  & 2.2 \\
%\hline
\textbf{Y-12}& 0.1  & 85 & 84.0 &0.55 &0.19   & 2.9 \\
%\hline
\textbf{Y-13}& 0.5  & $\sim 81$  &-- &-- &--  &3.5\\
\hline
\end{tabular}
\caption{Characteristic properties of the investigated samples; all the samples, except Y-00, have been irradiated at the thermal-neutron fluence of $1.476\times 10^{17}~\mathrm{cm^{-2}}$. $T_c^{on}$ is the temperature value at which $R_s/R_n = 1$; $T_{c0}$, $\sigma_{T_c}$ and $\lambda_0/\delta_0$ are the values used to obtain the best-fit curves of Fig.~\ref{fig:Rs(T)FIT}; $R_n$ is the normal-state value of the surface resistance obtained from Eq.~(1) at $T=T_c^{on}$.}
\label{tab:samples}
\end{table}

\subsection{Field-induced variation of Rs}

\begin{figure}[h!]
\centering
\includegraphics[width=0.45\textwidth]{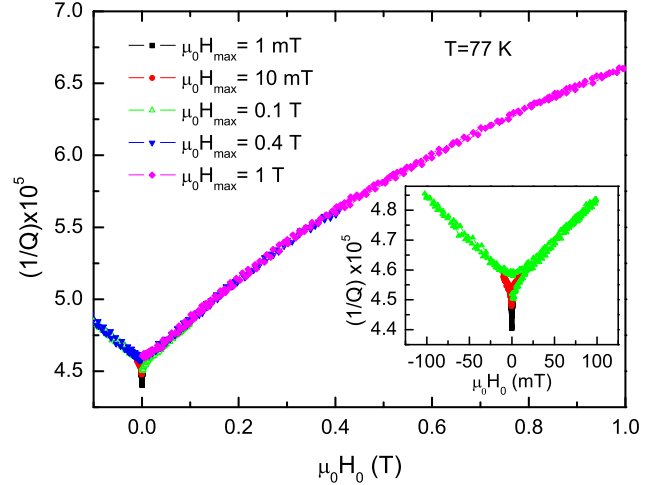}
\caption{Field-induced variations of 1/Q of the resonant cavity as a function of the DC magnetic field for the ZFC Y-00 sample, obtained sweeping $H_0$ in different ranges. The inset shows the results obtained during the first three field scans.}
\label{fig:Y00_77_1Q}
\end{figure}

The field-induced variations of $R_s$ have been investigated at different temperatures. For each measurement, the sample was zero-field cooled (ZFC) down to the desired value of temperature; the DC magnetic field was increased up to a certain value, $H_{max}$, and, successively, decreased down to zero. As an example of the typical behavior, Fig.~\ref{fig:Y00_77_1Q} shows the variation of $1/Q$ of the resonant cavity induced by the DC magnetic field, obtained in the sample Y-00 at $T=77$~K. Since the quality factor of the empty cavity does not depend on $H_0$, the field-induced variations of $1/Q$ are proportional to the field-induced variations of $R_s$. In particular, Fig.~\ref{fig:Y00_77_1Q} has been obtained sweeping $H_0$ in successive wider and wider ranges; the inset shows a zoom of the first three scans obtained sweeping the field from $-H_{max}$ to $+H_{max}$ with $\mu_0 H_{max}= 1~\mathrm{mT},~10~\mathrm{mT}$ and 0.1~T. As one can see, we observe a rapid variation at low fields followed by a slower one. 

In single crystals, the DC field value at which $R_s$ starts to increase coincides with the first penetration field of Abrikosov fluxons; on the contrary, in polycrystalline samples field variations can occur at lower fields because of the Josephson-fluxon penetration in weak links. The initial rapid variation of $R_s$ of Fig.~\ref{fig:Y00_77_1Q} can be reasonably ascribed to weak-link effects; this is corroborated by the presence of the magnetic hysteresis observed until $H_{max}$ reaches a certain value. Above a certain threshold value of $H_{max}$ (0.1~T in the figure), the $R_s(H)$ curve becomes reversible and the initial variation disappears irreversibly because the trapped magnetic flux very likely decouples the weak links. 

Results qualitatively similar as those of Fig.~\ref{fig:Y00_77_1Q} have been observed in all the investigated samples, except at temperatures very near $T_c$, where the low-field hysteresis is not present. Hereafter, first we discuss the results obtained in the pristine sample and, subsequently, those relative to other samples.  

\begin{figure}[h!]
\centering \includegraphics[width=0.45\textwidth]{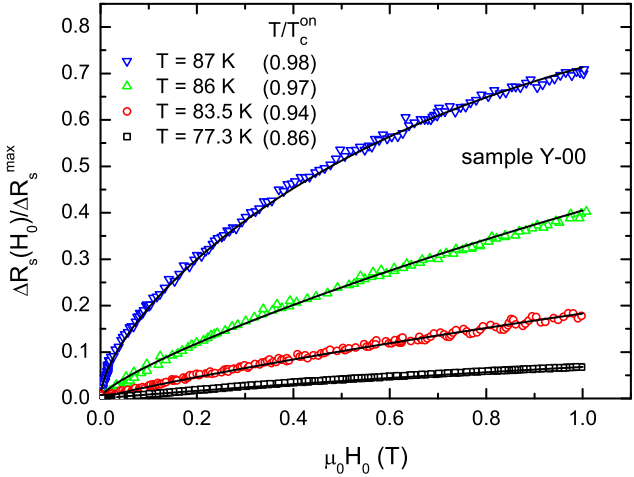}
\caption{Field-induced variations of $R_s$ at different temperatures for sample Y-00. $\Delta R_s(H_0)\equiv R_s(H_0,T)-R_s(0,T)$ and $\Delta R_s^{max}\equiv R_{n}-R_s(0,T)$. The lines are the best-fit curves obtained, as explained in the text, using $\mu_0 H_{c2}(0)=310$~T and $\gamma=5.8$ for all the curves; the values of $\omega_0/\omega$ are 0 for the three curves at $T=87$~K, 86~K, 83.5~K and 0.7 for the curve obtained at $T=77.3$~K.} \label{Fig:Rs(H)Y00}
\end{figure}

In Fig.~\ref{Fig:Rs(H)Y00} we report the field induced variations of the mw surface resistance for sample Y-00 at different values of the temperature (symbols), the continuous lines are the best-fit curves obtained as explained in the following. In the figure, $\Delta R_s(H_0)\equiv R_s(H_0,T)-R_s(0,T)$; moreover, the data are normalized to $\Delta R_s^{max}\equiv R_{n}-R_s(0,T)$. The experimental curves reported in the figure refer to the widest sweep, after the sample had been exposed to high magnetic field and, therefore, when the contribution of weak links is taken out.  

The complex penetration depth of the em field in superconductors in the mixed state has been calculated in different approximations~\cite{CC,BRANDT}. Coffey and Clem (CC) have elaborated a comprehensive theory, in the framework of the two-fluid model of superconductivity~\cite{CC}, under the assumption that the induction field, $B$, is uniform within the AC-field penetration depth; so, it is valid for applied fields higher enough than the first-penetration field whenever the effects of the critical state can be neglected~\cite{noistatocritico}. On the other hand, we do not observe magnetic hysteresis except in the field range in which the field-induced mw losses are mainly due to the presence of weak links. This can be ascribed to the fact that in the range of temperatures investigated $J_c$ is small enough that the effects of the critical state can be neglected.

In the linear approximation, $H_{\omega} \ll\ H_0$, the complex penetration depth expected from the CC model is given by
\begin{equation}\label{lambdatildeB}
    \widetilde{\lambda}(\omega,B,T)=\sqrt{\frac{\lambda^{2}(B,T)+
    (i/2)\widetilde{\delta}_{v}^{2}(\omega,B,T)}
    {1-2i\lambda^{2}(B,T)/\delta _{nf}^{2}(\omega,B,T)}}\, ,
\end{equation}
where $\lambda$ and $\delta_{nf}$ depend on $B$ because of the excitation of quasiparticles by the DC magnetic field, and $\widetilde{\delta} _{v}$ is the effective complex skin depth arising from the vortex motion.

$\widetilde{\delta} _{v}$ depends on the relative magnitude of the viscous-drag and the restoring-pinning forces. An important parameter that determines the regime of the fluxon motion is the so-called depinning frequency, $\omega_0$~\cite{gittle}. When $\omega \ll \omega_0$, the fluxon motion is ruled by the restoring-pinning force. On the contrary, for $\omega \gg \omega_0$, the contribution of the viscous-drag force predominates and the induced em current makes fluxons move in the flux-flow regime, with enhanced mw energy losses. In this case, $\widetilde{\delta} _{v}= \delta_{f}$ that, in the Bardeen-Stephen model~\cite{BS}, is given by $\delta_{f}= \delta_0 \sqrt{B/B_{c2}}$. In the more general case, one has
\begin{equation}\label{delta-v(omega)}
\frac{1}{\widetilde{\delta}_{v}^{2}}=\frac{1}{\delta_{f}^{2}}\left(1+
i~\frac{\omega_0}{\omega}\right)\,.
\end{equation}

Concerning the magnetic-field dependence of $\lambda$ and $\delta_{nf}$, one has to consider the excitation of quasiparticles induced by the applied field. This is expected to follow different laws in conventional (s-wave) and in d-wave superconductors. On the other hand, several experimental and theoretical investigations on HTS have clearly indicated the presence of a line of zeros in the superconducting order parameter consistent with the d-wave pairing symmetry~\cite{calorespecifico,harlingen,sigrist,Won}. General expressions of $\lambda(B,T)$ and $\delta_{nf}(\omega,B,T)$ can be written as
\begin{equation}\label{lamdaB}
\lambda^2(B,T) = \frac{\lambda_0^2}{[1-w_0(T)] \{1-[B /B_{c2}(T)]^{\alpha}\}}\,,
\end{equation}

\begin{equation}\label{deltaB}
\delta_{nf}^2(\omega,B,T) = \frac{\delta_0^2(\omega)}{1-[1-w_0(T)]\{1- [B /B_{c2}(T)]^{\alpha}\}}\,,
\end{equation}
where $1- [B /B_{c2}(T)]^{\alpha}$ accounts for the quasiparticle excitation induced by the magnetic field.\\ 
In conventional superconductors, where the quasiparticle density of state (DOS) comes from the low-energy states localized in the vortex cores, $\alpha=1$; in superconductors with lines of gap nodes, the DOS comes mostly from the continuous spectrum concentrated in proximity of the gap nodes outside the vortex core and, consistently with the calculations of Volovik~\cite{Volovik}, one can assume $\alpha=1/2$ for magnetic fields not close to $H_{c2}$. This assumption has been used to account for experimental results of field-induced mw losses in cuprate superconductors~\cite{Mallozzi,Silva_JSup}. Contrary to what occurs in conventional superconductors, in which the magnetic-field-induced mw losses are mainly due to the fluxon motion, it has been shown that in HTS an important contribution comes from the field-induced pair breaking, related to the gap nodes and described by a $\sqrt{B}$ dependence.\\

At fixed temperature, the expected field-induced variations of $R_s/R_n$ depend, besides the parameters already determined by fitting the data at $H_0=0$, on $\omega_0$, $H_{c2}(T)$ and the first penetration field, $H_p$. Moreover, one has to consider the anisotropy of the critical fields because the different orientations of the crystallites in the sample play an important role in determining the magnetic-field variations of $R_s$~\cite{fricano}. $H_p$ can be estimated from the decreasing-field branch of $R_s(H_0)$ measuring the field value at which $R_s$ does not change anymore; however, it does not play an important role because in the temperature range investigated is small and furthermore the model is valid for $H_0>2H_p$, where $B$ can be considered uniform over the sample. Since $H_{c2}$ is much larger than the maximum field we can achieve, except at temperatures very near $T_c$, we have measured $\partial H_{c2}(T)/\partial T|_{T_c}$ and estimated $H_{c2}(0)$, using the WHH formula~\cite{WHH}. Moreover, we have assumed:
\begin{equation}\label{Hc2(T)}
H_{c2}(T)=H_{c20}[1-(T/T_c)^2]\,.
\end{equation}
The $H_{c2}(T)$ values so obtained coincide with the upper critical field of the crystallites oriented with the \emph{c}-axis normal to the DC field. In order to take into account the anisotropy, we have assumed that the polycrystalline sample is constituted by grains with the $c$-axis randomly oriented with respect to the DC-magnetic-field direction; so, the distribution of their orientations follows a $\sin(\theta)$ law, being $\theta$ the angle between $\emph{\textbf{H}}_0$ and $\emph{\textbf{c}}$. Furthermore, we have used for the angular dependence of the upper critical field the anisotropic Ginzburg-Landau relation
\begin{equation}
    H_{c2}(\theta) = \frac{H_{c2}^{\perp
    c}}{\sqrt{\gamma^2 \cos^2(\theta) + \sin^2(\theta)}} \,,
\end{equation}
where $\gamma = H_{c2}^{\perp c}/H_{c2}^{\parallel c}$ is the anisotropy factor. 

Also the value of $H_p$ we estimate experimentally coincides with the first penetration field of crystallites orientated with the \emph{c}-axis normal to the DC field; therefore, its dependence on $\theta$ follows the same law of $H_{c1}(\theta)$ that, in the framework of the Ginziburg-Landau theory, is given by
\begin{equation}
    H_{c1}(\theta) = \frac{H_{c1}^{\parallel
    c}}{\sqrt{\cos^2(\theta) + \gamma^2 \sin^2(\theta)}} \,.
\end{equation}

Finally, we have used the following approximate expression for the induction field
\begin{equation}\label{B}
    B=\mu_0 \left(1+\frac{H_p}{H_{c2}-H_p}\right)(H_0-H_p)\,.
\end{equation}

In summary, the free parameters to fit the $R_s(H_0)$ curves at fixed temperature are $\omega_0$ and $\gamma$. Furthermore, it is necessary to find $H_{c20}$, letting it vary within the experimental uncertainty, because the measured value of $\partial H_{c2}(T)/\partial T|_{T_c}$ is determined with an experimental uncertainty of about 20\%. 

To fit our data, we have tried to use Eqs.~(\ref{lamdaB}) and (\ref{deltaB}) with $\alpha=1$, but this attempt did not give good fits, no matter the parameters used; also, by disregarding the anisotropy we did not get good results. For this reasons, we have set $\alpha=1/2$ and averaged the expected curves using the same distribution function of $T_c$ discussed in Section~\emph{3.1} and the $\sin(\theta)$ distribution function for the orientation of the crystallites with respect to the DC magnetic field. However, we would like to remark that the averaging procedure for taking into account the distribution of $T_c$ over the sample does not affect the theoretical results except at temperatures very near $T_c$.   

In order to fit the experimental data of Fig.~\ref{Fig:Rs(H)Y00}, we have used the following procedure. Firstly, we have fitted the results obtained at $T=87$~K, taking $H_{c20}^{\perp c}$ and $\gamma$ as parameters and setting $\omega_0=0$ in Eq.~(\ref{delta-v(omega)}); this is reasonable because at temperatures very near $T_c$ one can suppose fluxons move in the flux-flow regime. In this way, we have obtained $\gamma=5.8\pm 0.1$ and $\mu_0 H_{c20}^{\perp c}=310\pm 10$~T, consistently with the values reported in the literature for YBCO. Maintaining the same values of $H_{c20}^{\perp c}$ and $\gamma$, we have fitted the data at the other temperatures taking $\omega_0/\omega$ as parameter. The best-fit curves reported in Fig.~\ref{Fig:Rs(H)Y00} have been obtained with $\omega_0/\omega=0$ for $T=87$~K, 86~K and 83.5~K, and $\omega_0/\omega=0.7$ for $T=77.3$~K. As one can see, the experimental results obtained in sample Y-00 are quite well justified in the framework of Coffey and Clem theory provided that the anisotropy of the critical fields is taken into the due account and that the excitation of quasiparticles induced by the applied field is assumed to depend on $\sqrt{B}$, according with the d-wave model for cuprate superconductors. The analysis of the results of Fig.~\ref{Fig:Rs(H)Y00} shows that in a range of temperature of at least six degrees below $T_c$ fluxons move in the flux-flow regime; i.e., the restoring-pinning force is ineffective to hinder the fluxon motion induced by the mw current. At $T=77.3$~K, we obtain $\omega_0/2\pi\approx 6.7$~GHz consistent with the values reported in the literature~\cite{GOLOSOVSKY,SilvaReview,Tsuchiya}, though these studies have been done mainly in YBCO films and crystals.     

\begin{figure}[h!]
\centering \includegraphics[width=0.45\textwidth]{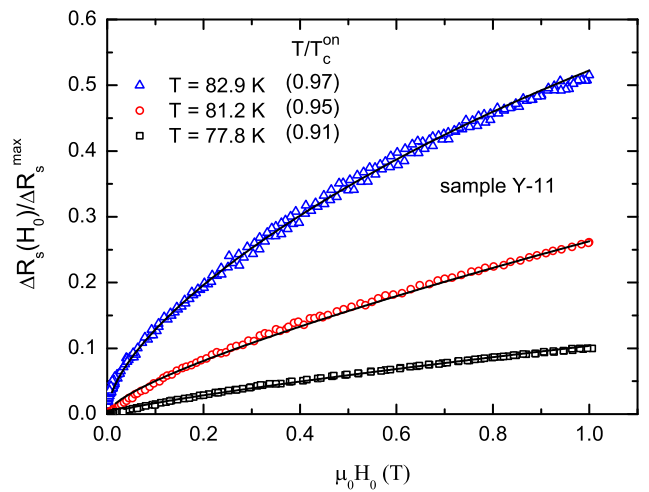}
\caption{Field-induced variations of $R_s$ at different temperatures for sample Y-11. The lines are the best-fit curves obtained, as explained in the text, using $\mu_0 H_{c2}(0)=310$~T and $\gamma=5.4$ for all the curves; the values of $\omega_0/\omega$ are 0 for the curve at $T=82.9$~K and $T= 81.2$~K, 0.5 for the curve obtained at $T=77.8$~K.} \label{Fig:Rs(H)Y11}
\end{figure}

The procedure used to fit the $R_s(H_0)$ curves of sample Y-00 has been repeated to fit the experimental data obtained in the samples Y-11 and Y-12. In Fig.~\ref{Fig:Rs(H)Y11}, we report the results obtained in sample Y-11 at different temperatures; symbols are the experimental data, continuous lines are the best-fit curves. By fitting the data at $T=82.9$~K, setting $\omega_0=0$, we have obtained $\gamma=5.4\pm 0.1$ and $\mu_0 H_{c20}^{\perp c}=310\pm 10$~T. The best-fit curves at $T=81.2$~K and $T=77.8$~K have been obtained maintaining the same values of $\mu_0 H_{c20}^{\perp c}$ and $\gamma$ and using $\omega_0/\omega$ as parameter. The values of $\omega_0/\omega$ that best fit the experimental data are $\omega_0/\omega=0$ for $T=81.2$~K and $\omega_0/\omega=0.5$ for $T=77.8$~K.   

\begin{figure}[t!]
\centering \includegraphics[width=0.45\textwidth]{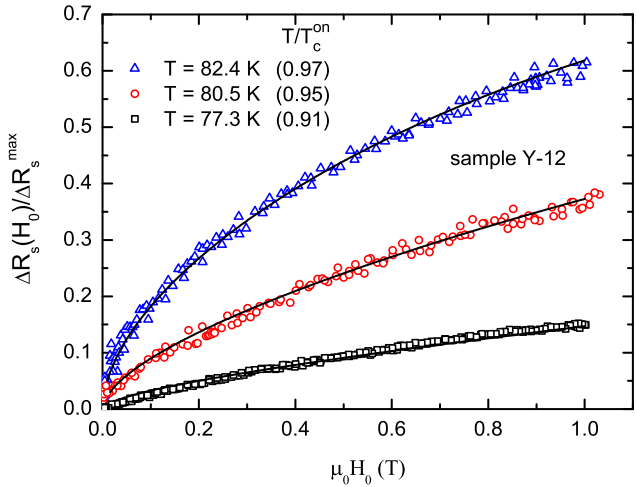}
\caption{Field-induced variations of $R_s$ at different temperatures for sample Y-12. The lines are the best-fit curves obtained, as explained in the text, using $\mu_0 H_{c2}(0)=250$~T and $\gamma=5$ for all the curves; the values of $\omega_0/\omega$ are 0 for the curve at $T=82.47$~K, 0.6 for the curve obtained at $T=80.5$~K and 1.5 for the that at $T=77.3$~K.} \label{Fig:Rs(H)Y12}
\end{figure}

Fig.~\ref{Fig:Rs(H)Y12} shows the field-induced variations of the mw surface resistance for sample Y-12 at different values of the temperature (symbols), the continuous lines are the best-fit curves obtained using the same procedure as samples Y-00 and Y-11. For this sample, by fitting the data at $T=82.4$~K, setting $\omega_0=0$, we have obtained $\gamma=5.0\pm 0.1$ and $\mu_0 H_{c20}^{\perp c}=250\pm 10$~T. The best-fit curves at lower $T$ have been obtained with the same values of $\gamma$ and $\mu_0 H_{c20}^{\perp c}$ and with $\omega_0/\omega= 0.6$ for $T=80.5$~K and $\omega_0/\omega=1.5$ for $T=77.3$~K.

By looking at the curves obtained in the three samples at $T/T_c=0.97$, in which the flux-flow regime has been hypothesized, one can note that the normalized values of the field-induced variations of $R_s$ increase on increasing the \BO\ content; this is essentially due to the higher values of $\lambda_0/\delta_0$ that, in turn, determine an increase of the mw-field penetration depth and enhance the field-induced pair-breaking effects. The effects of $\lambda_0/\delta_0$ on the expected results has been verified by numeric calculation.

The analysis of the experimental $R_s(H_0)$ curves highlights that the range of the reduced temperature in which fluxons move in the flux-flow regime shrinks on increasing the \BO\ addition. In particular, we have obtained nearly the same depinning-frequency value in sample Y-00 at $T/T_c^{on}=0.86$, in sample Y-11 at $T/T_c^{on}=0.91$ and in sample Y-12 at $T/T_c^{on}=0.95$. Moreover, by comparing the results obtained in samples Y-11 and Y-12, which are reported at the same values of $T/T_c^{on}$, one can see that the depinning frequency at $T/T_c^{on}=0.91$ in sample Y-12 is three time larger than that obtained in sample Y-11. This confirms that the effectiveness of the defects induced by the combined effect of the thermal-neutron irradiation and B doping on pinning the fluxons increases on increasing the \BO\ addition. 

Our results show that samples Y-00 and Y-11 have the same upper critical field, this is consistent with the fact that they have nearly the same value of $R_n$ and, therefore, the same normal-state resistivity, $\rho$; it means that the defects created in the sample with the 0.05~wt\% of \BO\ addition are unable to affect neither the electron time scattering nor the carrier density. On the contrary, the increase of $R_n$ of Y-12 implies that the induced defects give rise to an increase of $\rho$ of about 10\%. In conventional SC, an increase of $\rho$ is related to an increase of $H_{c2}$ because on increasing the defect density the mean free path reduces affecting the coherence length and, consequently, $H_{c2}$. On the contrary, in HTS it has been found that on increasing the defect density, $\rho$ increases and $|\partial H_{c2}(T)/\partial T|_{T_c}$ decreases~\cite{Krasno}. This finding has been ascribed to a reduction of the charge carries due to an increase of the defect density. On the other hand, several authors~\cite{Stefanescu,Topal2005,Margiani,Katsura} have suggested that dopants as Li and B can substitute Cu atoms in the Cu-O chains, breaking their continuity and, consequently, reducing their ability to transfer holes to the superconducting CuO$_2$ planes. So, the $\rho$ increase associated to the $H_{c2}$ decrease, we obtained in Y-12, may be due to the defect-induced reduction of carrier density.

Another effect of the \BO\ addition is to reduce the anisotropy factor, which goes from 5.8 in the pristine sample to 5 in sample Y-12. This finding agrees with the observation that on increasing the defect concentration anisotropy reduces~\cite{SauPRB51,Tonies,PRB43_91,Sawh,Marinaro}; it may be ascribed to creation of point defects randomly distributed over the sample, and/or to change in the oxygen stoichiometry that is consistent with the $T_c$ reduction.

\section{Conclusion}
We have measured the microwave surface resistance of melt-textured \YBCO\ samples, doped with different amount of \BO\ and, subsequently, irradiated by thermal neutrons at the fluence of $1.476\times 10^{17}~\mathrm{cm^{-2}}$. The mw surface resistance, $R_s$, has been measured, at $H_0=0$, as a function of the temperature and, at fixed temperatures, as a function of the DC magnetic field. The experimental results have been discussed in the framework of the Coffey and Clem theory, properly adapted to take into account the d-wave nature of the cuprate superconductors. In particular, the excitation of quasiparticles induced by the temperature has been supposed to vary quadratically with the temperature and that induced by the magnetic field to follow the $\sqrt{B}$ law. We would like to remark that all the attempts done without the above conditions failed; this confirm that the d-wave nature of YBCO material strongly affects the mw energy losses. Moreover, it has been necessary to consider the anisotropy of the critical fields, and the distribution of $T_c$ over the samples due to the inhomogeneity of the polycrystalline samples.   

We have found that the defects induced by the combined effects of the neutron irradiation and the B doping affect both the temperature and magnetic-field dependencies of $R_s$. By fitting the $R_s(T)$ curves, at $H_0=0$, we have found that the main effect of the increase of the \BO\ content is that to enlarge the effective field penetration depth. By fitting the data of the magnetic-field-induced variations of $R_s$, we have determined the depinning frequency, at different temperatures, which is a gauge of the effectiveness of the defects to hinder the fluxon motion induced by the mw current. We have found that the depinning frequency increases on increasing the \BO\ addition, confirming the increase of the pinning strength. However, a high \BO\ content such as 0.5 wt\% degrades the material properties in both the superconducting and normal states.  

% \appendix

\section*{Acknowledgements}
The authors are very glad to thank M. Bonura for his interest to this work and helpful suggestions; G. Napoli for technical assistance.
% \label{}

% Bibliographic references with the natbib package:
% Parenthetical: \citep{Bai92} produces (Bailyn 1992).
% Textual: \citet{Bai95} produces Bailyn et al. (1995).
% An affix and part of a reference:
%   \citep[e.g.][Ch. 2]{Bar76}
%   produces (e.g. Barnes et al. 1976, Ch. 2).

\end{document}